\documentclass[twocolumn,showpacs,preprintnumbers,amsmath,amssymb]{revtex4}

\usepackage{graphicx}
\usepackage{dcolumn}
\usepackage{bm}
\usepackage{subfigure}

\begin{document}
\title{Band Structure Engineering of Multinary Chalcogenide Topological Insulators}

\author{Shiyou Chen$^{1,2}$, X. G. Gong$^2$, Chun-Gang Duan$^{1}$, Zi-Qiang Zhu$^{1}$, Jun-Hao Chu$^{1}$, Aron Walsh$^3$, Yu-Gui Yao$^4$, Jie Ma$^5$ and Su-Huai Wei$^5$}
\affiliation{$^1$Laboratory of Polar Materials and Devices, East
China Normal University, Shanghai 200241, China}
\affiliation{$^2$ Laboratory for Computational Physical Sciences and Surface Physics
Laboratory, Fudan University, Shanghai 200433, China}
\affiliation{$^3$ Department of Chemistry, University College London, London WC1E 6BT, UK}
\affiliation{$^4$ Institute Of Physics, Chinese Academy of Sciences, Beijing 100190, China}
\affiliation{$^5$ National Renewable Energy Laboratory, Golden, CO 80401, USA}

\date{\today}

\begin{abstract}
Topological insulators (TIs) have been found in strained binary HgTe
and ternary I-III-VI$_2$ chalcopyrite compounds such as CuTlSe$_2$
which have inverted band structures. However, the non-trivial band
gaps of these existing binary and ternary TIs are limited
to small values, usually around 10 meV or less. In this work, we
reveal that a large non-trivial band gap requires the material
having a large negative crystal field splitting $\Delta_{CF}$ at top
of the valence band and a moderately large negative $s-p$ band gap
$E_g^{s-p}$. These parameters can be better tuned through chemical
ordering in multinary compounds. Based on this understanding, we
show that a series of quaternary I$_2$-II-IV-VI$_4$ compounds,
including Cu$_2$HgPbSe$_4$, Cu$_2$CdPbSe$_4$, Ag$_2$HgPbSe$_4$ and
Ag$_2$CdPbTe$_4$ are TIs, in which Ag$_2$HgPbSe$_4$ has the largest
TI band gap of 47 meV because it combines the optimal values of
$\Delta_{CF}$ and $E_g^{s-p}$.
\end{abstract}

\maketitle

The search for new topological insulators (TIs) has intensified
recently due to their scientific importance as a novel quantum state
and the associated technological applications in spintronics and
quantum computing\cite{qi-33-2010,moore-194-2010}. So far,
experimental realizations have been limited to a few classes of
simple materials, including zinc-blende based HgTe quantum
wells\cite{konig-766-2007,bernevig-1757-2006,luo-176805-2010},
Bi$_{1-x}$Sb$_x$ alloys\cite{fu-045302-2007,hsieh-970-2008} and
binary tetradymite semiconductors such as Bi$_2$Se$_3$ and
Bi$_2$Te$_3$\cite{zhang-438-2009,xia-398-2009,chen-178-2009}. Most
recently, the search for TIs has extended to ternary
compounds\cite{xiao-096404-2010,chadov-541-2010,lin-546-2010,lin-036404-2010},
e.g., strained Half-Heusler compounds, in the hope that the presence
of more chemical elements would bring greater material flexibility.
Despite the success of identifying these TIs, the design of new TI
materials with the following advantages is still desired: (i)
realizing a topological insulating state with a significant
non-trivial band gap (i.e., larger than $k$T at room temperature) at
its natural equilibrium state (i.e., not under external strain),
(ii) easy integration with electronic and spintronic devices based
on tetrahedral semiconductors, and (iii) easy to be synthesized or
already have been synthesized.

Based on the direct evaluation of the $Z_{2}$ topological
invariant, Feng et al.\cite{feng-016402-2011} proposed that a series
of I-III-VI$_2$ chalcopyrite compounds (such as CuTlSe$_2$) could
have topologically non-trivial band structure, and some of them can
realize a topological insulating phase in their natural equilibrium structure.
This is an important observation because the chalcopyrite structure
is derived from the zinc-blende structure, and the band structure
properties are well understood, mostly for solar cell
applications\cite{madelung-2004,wei-3846-1995}. Some of the proposed
Cu and Ag based TIs, such as CuTlSe$_2$ and AgTlTe$_2$, have already
been synthesized
experimentally\cite{madelung-2004,landolt-borstein}. However, the
predicted band gaps of these TIs are very small, usually around 10
meV or less, similar to that observed in strained
HgTe\cite{bernevig-1757-2006}.

In this Letter, we show that the non-trivial band gaps of
zinc-blende derived compounds with inverted band structure are
mainly determined by the crystal field splitting $\Delta_{CF}$ at
the top of the valence band and the size of the inverted s-p band
gap $E_g^{s-p}$, which can be better tuned by changing the component
elements in a multinary ordered compounds. A large non-trivial band
gap requires the material having a large negative $\Delta_{CF}$ and a
large negative $E_g^{s-p}$ as long as it has no band crossing at the Fermi energy. For
I-III-VI$_2$ topological insulators, because the band inversion
requires the group-III elements to be large and heavy, whereas a
large negative $\Delta_{CF}$ requires group-III elements to be small
and light, the possibilities for obtaining a large TI band gap  are
limited. Through further cation mutation\cite{calculationmethods},
large negative $\Delta_{CF}$ and $E_g^{s-p}$ is achievable in
quaternary II$_2$-II-IV-VI$_4$ compounds. We have identified four
topological insulators (Cu$_2$HgPbSe$_4$, Cu$_2$CdPbSe$_4$,
Ag$_2$HgPbSe$_4$ and Ag$_2$CdPbTe$_4$), in which Ag$_2$HgPbSe$_4$
has the largest TI band gap of 47 meV. In the following, we will
discuss the evolution of the band structure of zinc-blende derived
structures and explain what kind of band structure can lead to the
largest TI band gap.

For a normal zinc-blende semiconductors such as CdTe, the band gap
is between the s-like conduction band minimum (CBM) $\Gamma_{6c}$
state and the p-like valence band maximum (VBM) $\Gamma_{8v}$ state,
as shown in Fig. 1.  The non-trivial band structure of a TI is
characterized by the band inversion in the Brillouin
zone\cite{fu-045302-2007,feng-016402-2011}, i.e., the position
of the conduction and valence bands is switched. In
zinc-blende compounds, the band inversion means that the
$\Gamma_{6c}$ level falls below the $\Gamma_{8v}$ level. In the
inverted band structure, the $\Gamma_{6c}$ level is occupied, while
the quadruply-degenerate $\Gamma_{8v}$ level is half occupied,
making the Fermi level stay at the $\Gamma_{8v}$ level and the
system become a zero-gap semi-metal. This is the case for bulk HgTe.

\begin{figure}
\scalebox{0.7}{\includegraphics{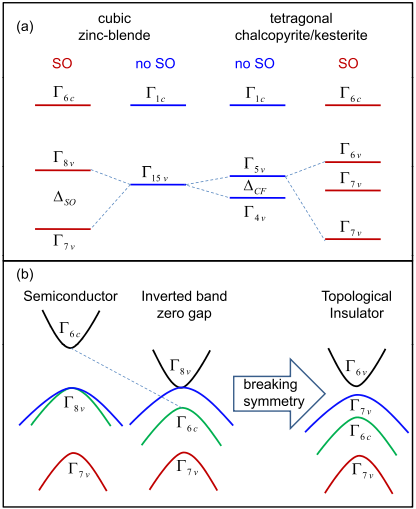}}
\caption{\label{fig1}(Color online) (a) The conduction and valence
band splitting of cubic and tetragonal semiconductors. (b) A plot
showing how the band structure of normal semiconductors transfers
into the inverted and topological insulator band structures. Note
that the subscript $v$ ($c$) represents the state belongs to the
valence (conduction) band in the normal band structure. }
\end{figure}

\begin{figure*}
\scalebox{1.0}{\includegraphics{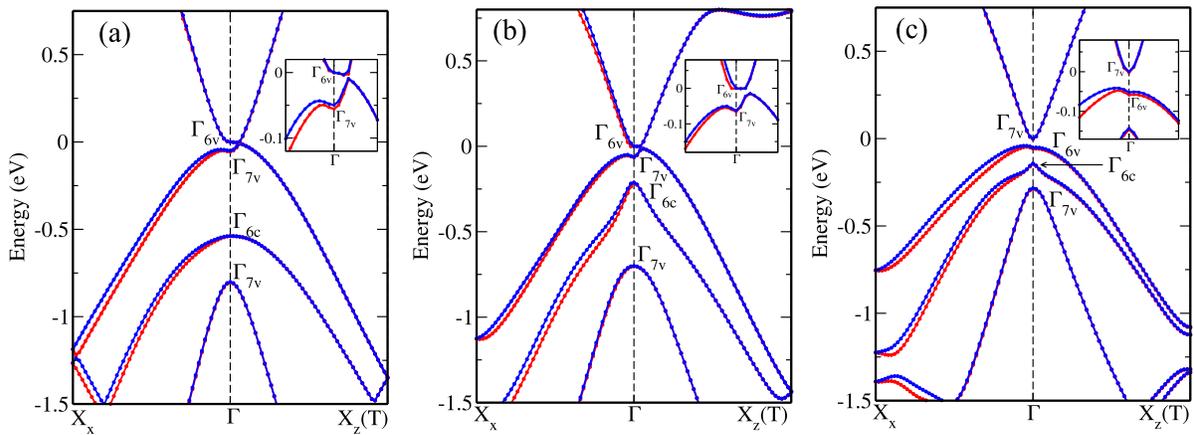}}
\caption{\label{fig2}(Color online) The calculated band structure
along the high symmetry lines, $X_X:\frac{2\pi}{a}(1~0~0)\rightarrow
\Gamma:(0~0~0) \rightarrow X_Z:\frac{2\pi}{a}(0~0~1)$  of (a) HgTe
with a (001) tensile strain and $\Delta_{CF}=70$ meV, (b) CuTlTe$_2$
and (c) Ag$_2$HgPbSe$_4$ at their equilibrium states. $X_X$ and
$X_Z$ are the notations of zinc-blende structure, and $X_Z$
corresponds to T in the chalcopyrite structure. Red and blue color
are used to show the two spin-dependent bands clearly.}
\end{figure*}

To open a band gap and change the zinc-blende semi-metal HgTe into a topological
insulator, one has to induce a crystal field splitting $\Delta_{CF}$
by reducing the $T_d$ symmetry of the zinc-blende structure to, e.g.,
$D_{2d}$, by applying an epitaxial strain or forming a quantum
well\cite{bernevig-1757-2006}. For $D_{2d}$ symmetry, the
half-filled $\Gamma_{8v}$ state splits into $\Gamma_{6v}$ and
$\Gamma_{7v}$ state, and a gap can be opened around the occupied
$\Gamma_{7v}$ ($\Gamma_{6v}$) and unoccupied $\Gamma_{6v}$
($\Gamma_{7v}$) levels (Fig. 1(b)) if $\Delta_{CF}$ is positive
(negative). On the other hand, the crystal field splitting can also
be induced by chemical ordering, e.g., by mutating two Hg (group II)
atoms into one Cu (group I) and one Tl (group III), forming ordered
I-III-VI$_2$ chalcopyrite compounds such as
CuTlTe$_2$\cite{feng-016402-2011,chen-165211-2009}.

In Fig. 2(a) we plot the calculated band structure of HgTe under a
$\epsilon=0.02$ (001) tensile strain with $\Delta_{CF}$=70 meV and
CuTlTe$_2$ in the chalcopyrite structure with $\Delta_{CF}$=76 meV.
As we can see, a small gap is opened near the $\Gamma$ point for
both systems. Although the size is small, this anticrossing gap is protected by
the lattice symmetry\cite{moon-045205-2006}. For the band structure calculation we
employed density functional theory with a hybrid
exchange-correlation functional, which can correctly predict the
band gaps of many zinc-blende and chalcopyrite
semiconductors\cite{paier-154709-2006,hummer-195211-2007,hse}.

Comparing the band structure of HgTe under a $\epsilon=0.02$ (001)
tensile strain and CuTlTe$_2$, we find that the overall shape is
very similar, especially near the band gap. In both systems, the
s-like $\Gamma_{6c}$ state falls below the p-like $\Gamma_{6v}$ and
$\Gamma_{7v}$ states, and the minimum gap occurs along the
$\Gamma-X_Z$ line. This similarity between strained HgTe and
CuTlTe$_2$ indicates that the strain and chemical ordering have the
same effect in producing the crystal field splitting $\Delta_{CF}$
at the top of valence band\cite{wei-14337-1994}, therefore, it could
be an efficient way to tune the TI band gap.

To achieve this goal, it is important to understand first how the
splitting at the top of valence band is influenced by chemical
ordering and what is the resulting dependence of the TI band gap.
Based on the quasi-cubic model\cite{rowe-451-1971,wei-14337-1994},
and assuming the $\Gamma_{6c}$ state is far away from the band edge,
we know that the splitting of the $\Gamma_{8v}$ level into
$\Gamma_{6v}$ and $\Gamma_{7v}$ under the tetragonal symmetry
depends on two quantities: the spin-orbit splitting $\Delta_{SO}$
and the crystal field splitting $\Delta_{CF}$. $\Delta_{CF}$ is
defined to be positive if the doubly-degenerate $\Gamma_{5v}$ is
above the singly-degenerate $\Gamma_{4v}$ state when the spin-orbit
interaction is not considered, as shown in Fig. 1(a). For systems
where $\Delta_{SO}$ is much larger than $\Delta_{CF}$, the splitting
between $\Gamma_{6v}$ and $\Gamma_{7v}$ is close to 2/3 of
$\Delta_{CF}$. Previous studies\cite{moon-045205-2006} on strained
zinc-blende compound showed that the non-trivial gap depends on the
sign and size of $\Delta_{CF}$: (i) when $\Delta_{CF} < 0$ the gap
occurs along of the $\Gamma-X_X$ line near the $\Gamma$ point and
the gap increases quickly as a function of the magnitude of
$\Delta_{CF}$; (ii) when $\Delta_{CF} > 0$, the gap occurs along the
$\Gamma-X_Z$ line near the $\Gamma$ point and the gap increases
slowly as a function of $\Delta_{CF}$. This can be seen clearly in
Fig. 3(a), where the dependence of the non-trivial band gap on the
size of $\Delta_{CF}$ for HgTe is plotted. For example, when
$\Delta_{CF}$=-100 meV, the gap is almost 40 meV, but when
$\Delta_{CF}$=100 meV, the gap is only 5 meV. The reason for the
more significant gap increase with negative $\Delta_{CF}$ is that,
$\Gamma-X_X$ line has lower symmetry than $\Gamma-X_Z$ line, so the
band anticrossing is more significant when the gap shifts to the
$\Gamma-X_X$ line. CuTlTe$_2$ has a calculated $\Delta_{CF}= 76$
meV. This positive value explains why the gap shifts to the
$\Gamma-X_Z$ line with only a small value of about 14 meV [Fig.
2(b)]. Based on this observation, we know that large gap TI can only
exist in zinc-blende derived compounds with large negative
$\Delta_{CF}$.

\begin{figure}
\scalebox{1.0}{\includegraphics{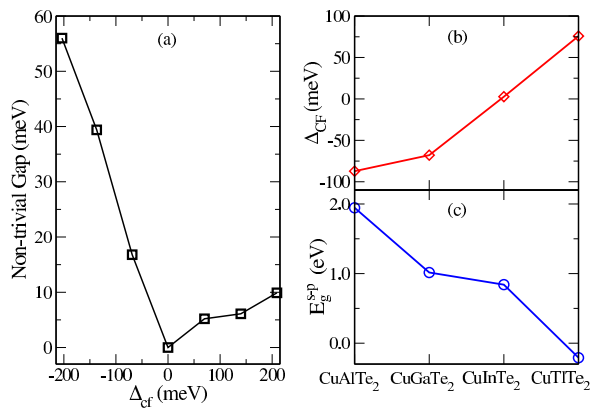}}
\caption{\label{fig3}(Color online) (a) The calculated non-trivial
band gap as a function of $\Delta_{CF}$ for HgTe. Here $\Delta_{CF}$
is changed by tuning the (001) strain $\epsilon$. (b) The calculated
$\Delta_{CF}$ and (c) $E_g^{s-p}$ of Cu-III-Te$_2$ with III=Al, Ga,
In, Tl.}
\end{figure}

In Fig. 3(b) we plot the calculated $\Delta_{CF}$ of CuAlTe$_2$,
CuGaTe$_2$, CuInTe$_2$ and CuTlTe$_2$. As we can see, $\Delta_{CF}$
increases from negative to positive as the group-III cations change
from Al to Tl, i.e., from small light to large heavy elements.
Considering that large negative $\Delta_{CF}$ enlarges the
non-trivial gap, one may intend to search compounds with small
group-III cation as candidates for TIs. However, the requirement of
band inversion at the $\Gamma$ point excludes Al, Ga and In
compounds because CuAlTe$_2$, CuGaTe$_2$ and CuInTe$_2$ all have the
normal band order. Their band gaps $E_g^{s-p}= E(\Gamma_{6c}) -
E(\Gamma_{6,7v})$ are all positive and decrease from Al to Ga to In
compounds. This can be understood according to the band component of
I-III-VI$_2$ chalcopyrites: the $\Gamma_{6c}$ state has s-like
anti-bonding character localized on group-III cation and group-VI
anion, whereas $\Gamma_{7v}$ and $\Gamma_{6v}$ states mainly have
the $p$ component of the group-VI anion hybridized with the \emph{d}
component of the group-I
cation\cite{jaffe-5822-1983,chen-205209-2007}. Two factors shift the
$\Gamma_{6c}$ level down from Al to Ga to In compounds
\cite{wei-3846-1995,chen-205209-2007,chen-165211-2009}: (i) the
\emph{s} orbital energy of Ga is deeper than Al and (ii) In is much
larger than Ga. For Tl, its \emph{s} orbital energy, like Hg, is
very deep due to the large relativistic effect, so its band gap is
much lower than that of the corresponding In compounds. This is
confirmed in Fig. 3(c), where we plot the calculated $E_g^{s-p}$ of
CuAlTe$_2$, CuGaTe$_2$, CuInTe$_2$ and CuTlTe$_2$; only CuTlTe$_2$
has negative $E_g^{s-p}$, i.e., its \emph{s}-like $\Gamma_{6c}$
state falls below the p-like states at $\Gamma$ point (band
inversion). Unfortunately, CuTlTe$_2$ has a positive
$\Delta_{CF}=76$ meV, thus only a small TI band gap of about 14 meV.

In the above discussion, we have assumed that the $\Gamma_{6c}$
state is deep inside the valance band, thus has no effect on the
band splitting and the non-trivial gap of the TIs. However, if the
$\Gamma_{6c}$ is close to the band edge, then we have to consider
its interaction with the band edge states. This is because when
$\Delta_{CF} < 0$, the band gap of the TI at $\Gamma$ point is
between the unoccupied $\Gamma_{7v}$ and the occupied $\Gamma_{6v}$
(or $\Gamma_{6c}$, if it has a higher energy than $\Gamma_{6v}$)
derived state. The coupling between the $\Gamma_{6v}$ and
$\Gamma_{6c}$ states pushes the $\Gamma_{6v}$ level up in energy,
thus reduces the effective crystal field splitting between the
$\Gamma_{7v}$ and $\Gamma_{6v}$ state and the non-trivial band gap.
This is what we find for AgTlSe$_2$ and AgTlTe$_2$. According to our
calculation, the non-trivial gap of AgTlSe$_2$ is limited at
$\Gamma$ point with a very small size, 1 meV, although it has a
large negative $\Delta_{CF}= -50$ meV. Therefore, to reduce the
interaction between the $\Gamma_{6v}$ and $\Gamma_{6c}$ states, one
should move the $\Gamma_{6c}$ level down, i.e., increase the
magnitude of negative $E_g^{s-p}$ as much as possible.

The above analysis indicates that to design large gap chalcopyrite
I-III-VI$_2$ TIs, we face two contradictory requirements (large
negative $\Delta_{CF}$ and large negative $E_g^{s-p}$). This
severely limits the largest non-trivial gap obtainable for
I-III-VI$_2$ compounds. Through the direct calculation, we find that
most of the already-synthesized I-III-VI$_2$ have positive
$E_g^{s-p}$ and are normal semiconductors\cite{madelung-2004},
except CuTlSe$_2$, CuTlTe$_2$, AgTlSe$_2$ and AgTlTe$_2$. But the
non-trivial gaps of these four TIs are all small due to the positive
$\Delta_{CF}$ for CuTlSe$_2$ and CuTlTe$_2$, and small $E_g^{s-p}$
for AgTlSe$_2$ and AgTlTe$_2$.

To further increase the non-trivial band gap, we need to make both
the $\Delta_{CF}$ and $E_g^{s-p}$ more negative. We find that this
can be done by mutating two group-III cations in I-III-VI$_2$
compounds to one group-II and one group-IV cation, thus forming the
I$_2$-II-IV-VI$_4$ (I=Cu, Ag, II=Zn, Cd, Hg, IV=Si, Ge, Sn, Pb,
VI=S, Se, Te) quaternary compounds. These compounds crystallize in
either tetrahedral kesterite or stannite structures. Due to the
increased chemical and structural freedom in the quaternary
compounds, their band structure can be better tuned. Also, because
they are structurally derived from chalcopyrites, their band
structures maintain similar
characteristics\cite{chen-165211-2009,chen-041903-2009}. Therefore,
if these compounds have inverted band structure, they can also be
TIs.

Similar to the chalcopyrites, we need to have compounds that contain
heavy group-IV elements so that the $\Gamma_{6c}$ level could fall
below the $\Gamma_{6v}$ and $\Gamma_{7v}$
levels\cite{chen-165211-2009,chen-041903-2009}. Table I lists the
calculated $E_g^{s-p}$ of I$_2$-II-Pb-VI$_4$ compounds. The results
show that most of the Pb-Te and Pb-Se compounds have negative
$E_g^{s-p}$ at $\Gamma$ and are therefore candidates for TIs. The
calculation also shows all sulphides and compounds containing other
group-IV cations (Sn, Ge, Si) have positive $E_g^{s-p}$ and are
normal semiconductors.

\begin{table}
\caption{\label{tab:table1} The calculated $E_g^{s-p}$ of
I$_2$-II-Pb-VI$_4$ (I=Cu, Ag, II=Cd, Hg, VI=S, Se, Te) in their
ground-state structure. TM, TI and NI in the parentheses represent
topological metal, topological insulator and normal insulator,
respectively.}
\begin{ruledtabular}
\begin{tabular}{lcccc}
            &   Structure        &  Te$_4$       &  Se$_4$      & S$_4$       \\
Cu$_2$HgPb  &   stannite         &    -0.46 (TM) &  -0.32 (TI)  &  0.07 (NI)  \\
Cu$_2$CdPb  &   stannite         &    -0.21 (TM) &  -0.07 (TI)  &  0.32 (NI)  \\
Ag$_2$HgPb  &   kesterite        &    -0.37 (TM) &  -0.14 (TI)  &  0.40 (NI)  \\
Ag$_2$CdPb  &   kesterite        &    -0.12 (TI) &  0.18  (NI)  &  0.72 (NI)  \\
\end{tabular}
\end{ruledtabular}
\end{table}

We first look at the band structure of Cu$_2$HgPbTe$_4$ which has
the most negative $E_g^{s-p}$. The overall shape near the $\Gamma$
point is similar to those of ternary CuTlTe$_2$ and binary HgTe
under (001) strain as shown in Fig. 2(a) and 2(b), indicating that
the band structure character is kept in the cation mutation.
However, Cu$_2$HgPbTe$_4$ is actually a topological metal (TM),
because the conduction band near L(N):$\frac{2\pi}{a}(0.5~0.5~0.5)$
point drops below VBM and crosses the Fermi level. The reason is
that the conduction band state at L(N) point has similar character
to the $\Gamma_{6c}$ state, so when the $\Gamma_{6c}$ energy is too
low, the $L_{1c}(N_{1c})$ state energy is also below VBM, making the
system metallic. To avoid this situation, therefore, we should
search for TI material with mildly negative $E_g^{s-p}$.

Our previous study\cite{chen-205209-2007,chen-165211-2009} has shown
that replacing Cu by Ag or replacing Te by Se can increase
$E_g^{s-p}$, i.e., raising the $\Gamma_{6c}$ and $L_{1c}$ energy
level relative to $\Gamma_{6v}$ and $\Gamma_{7v}$, because (i) at
the top valence band, the lower \emph{4d} level and larger size of
Ag compared to Cu weakening the p-d hybridization, and the \emph{4p}
level of Se is lower than \emph{5p} level of Te, which both shift
the $\Gamma_{6v}$ and $\Gamma_{7v}$ levels down, (ii) the
displacement of anion towards Pb in the Ag compounds and the smaller
size of Se than Te also both reduce the Pb-anion bond lengths,
increasing the energy of the Pb(s)-anion(s) antibonding states at
the bottom conduction band. This expectation is supported by the
calculated band structure of Ag$_2$HgPbSe$_4$, which has no band
crossing at the Fermi level and thus is a topological insulator, as
shown in Fig. 2(c). Similarly, we predict that Cu$_2$CdPbTe$_4$ and
Ag$_2$HgPbTe$_4$ are topological metals, while Cu$_2$HgPbSe$_4$,
Cu$_2$CdPbSe$_4$ and Ag$_2$CdPbTe$_4$ are topological insulators.
The results are shown in Table I.

Among the four identified quaternary TIs, Ag$_2$HgPbSe$_4$ has the largest non-trivial gap of
47 meV. This is because Ag$_2$HgPbSe$_4$ is more stable in the low symmetry kesterite structure
with large group-I element, therefore, it has a large negative $\Delta_{CF}$ (-51 meV). It also
has a reasonably large negative $E_g^{s-p}$ gap, so the coupling between the $\Gamma_{6c}$ and
the $\Gamma_{6v}$ state is weak. Its band structure is shown in Fig. 2(c).
As expected, we see the gap of Ag$_2$HgPbSe$_4$ occurs at a position along
the low symmetry $\Gamma-X_X$ line, consistent with our discussion above.

In conclusion, we have shown that the non-trivial band gaps of
zinc-blende derived topological insulators depend on the crystal
field splitting at the top valence band as well as the size of the
inverted s-p band gap. In general, a material with large TI band gap
should have a large negative crystal field splitting and a moderate
size of the inverted band gap. Compared to binary zinc-blende and
ternary chalcopyrite compounds, these parameters can be more easily
tuned through the chemical ordering in quaternary compounds. Based
on this understanding, we have identified four ground state quaternary
topological insulators, among which Ag$_2$HgPbSe$_4$ has the largest
TI band gap of 47 meV because it has the optimal band structure
parameters.

This work is supported by NSF of Shanghai (No. 10ZR1408800) and China
(No. 10934002, 10950110324 and 10974231), the Research Program of
Shanghai municipality and MOE, the Special Funds for Major State
Basic Research, the Fundamental Research Funds for the Central
Universities, PCSIRT and 973 Program (No. 2007CB924900).  The work
at NREL is funded by the U.S. Department of Energy, under Contract
No. DE-AC36-08GO28308.


\end{document}